\documentclass[aps,pre,a4paper,10pt,notitlepage,superscriptaddress]{revtex4-1}
\usepackage[english]{babel}
\usepackage{lmodern}
\usepackage[latin9]{inputenc}
\usepackage{endnotes}
\usepackage{amssymb,amsmath,amsfonts}
\usepackage{textcomp}

\usepackage{graphicx}% Include figure files
\usepackage{dcolumn}% Align table columns on decimal point
\usepackage{bm}% bold math
\usepackage{xcolor}

\begin{document}

\title{Shearing effects on the phase coarsening of binary mixtures using the Active Model B}
\author{A. Lamura}
\affiliation{Istituto Applicazioni Calcolo CNR, via dei Taurini 19, Rome, Italy}
\author{A. Tiribocchi}
\affiliation{Istituto per le Applicazioni del Calcolo CNR, via dei Taurini 19, Rome, Italy}

%\date{\today}

\begin{abstract}
The phase separation of a two-dimensional active binary mixture is studied under the action of an applied shear through numerical simulations. 
It is highlighted how the strength of the external flow modifies the initial shape of growing domains. 
The activity is responsible for the formation of isolated droplets which affect both the coarsening dynamics and the morphology 
of the system. The characteristic dimensions of domains along the flow and the shear direction are modulated in time by oscillations whose amplitudes are reduced when the activity increases. This induces a broadening of the distribution functions of domain lengths 
with respect to the passive case due to the presence of dispersed droplets of different sizes.
\end{abstract}

\maketitle

\section{Introduction}
Over the last decades much research has been dedicated to studying active materials, i.e., non-equilibrium systems in which internal units continuously consume energy, usually stored in the environment, to self-propel~\cite{ramaswamy2010,roman2012,marchetti2013,bechinger2016,ramaswamy2019}. Examples of active matter range from bacterial and algal suspensions~\cite{dileonardo2010ratch,dogic2013,clement2015,arciprete2018} to colloidal particles acquiring motion through chemical reactions~\cite{therekauff2012,buttinoni2013,palacci2013}, up to the cytoskeleton of living cells~\cite{needleman2017,yeomans2017,shaebani2020}, to name but a few. Importantly, to convert their energy into motion, particles violate time-reversal symmetry (TRS) at the micro-scale.

An active process of particular relevance to us is the motility-induced phase separation (MIPS), where a suspension of repulsive motile particles can phase separate into bulk dense (liquid) and dilute (vapor) regions~\cite{tailleur2008,cates2015,gonnella2015}.  This essentially occurs because active particles aggregate where they move slowly, thus leading to a local density increase causing a slowing down and further accumulation. 
The resulting clusters show highly dynamical particle exchange leading to macroscopic phase separation~\cite{fily2012,stenhammar2013,stenhammar2014}.  The kinetics of MIPS has been found to share many features with that of passive systems with attractive interactions, thus suggesting that TRS could be restored at the macroscale level~\cite{tailleur2008, cates2015,speck2014,fodor2016,nardini2017,takatori2014,solon2015,ginot2015}. However, large scale simulations of active Brownian particles~\cite{stenhammar2013,stenhammar2014, tjhung2018,caporusso2020} as well as experiments~\cite{therekauff2012,buttinoni2013,palacci2013} reveal that MIPS may also {exhibit} further non-equilibrium features, such as the formation of mesoscopic vapor bubbles within the particles aggregates and microphase separation~\cite{stenhammar2013,tjhung2018,caporusso2020,shi2020}. This scenario would indeed support the view that time-reversal symmetry is manifestly broken macroscopically. 

Alongside particle-based simulations, the~phenomenology of MIPS can be also described using continuum field theories, either by an explicit coarse-graining of the microscopic dynamics or via symmetry arguments and conservation laws~\cite{cates2019}. The~latter strategy has led to the construction of the Active Model B (AMB) \cite{nardini2017,tjhung2018,witt2014}, a~field theory that, in~the absence of solvent, essentially extends the passive model B~\cite{modb} by incorporating an active gradient term
that cannot be obtained from a free-energy functional, thus
breaking time-reversal symmetry. 
The passive Model B (MB) falls within a class of phase-field models describing the kinetics of phase separation through a conserved scalar order parameter $\phi$ capturing, for~example, the~density of colloids or the concentration in a binary mixture. If~hydrodynamics can be neglected, the~evolution of $\phi$ is governed by a Cahn-Hilliard equation~\cite{ch1,ch2},  where the thermodynamic force driving the relaxation of $\phi$ stems from the functional derivative of a $\phi^4$ free energy with square gradient terms. Such simplified description captures, for~example, the~result $R\sim t^{\alpha}$, for~the dependence of the domain size $R$ on time $t$ with a characteristic growth exponent $\alpha=1/3$ \cite{BRAY}.

MB provides a robust framework to describe the passive phase separation subject to an external shearing {\cite{doi1990,lamu1998-1,lamu1998-2,noisim-1,noisim-2,berthier2001,lamu2007}}. In~that case the growth is anisotropic, being characterized by domains
stretched and tilted along the flow direction as observed in simulations~\cite{rothman,noisim-2}
and experiments~\cite{hashi}. 
Indeed, numerical solutions of the dynamical equation with shear~\cite{noisim-2} confirm the presence of anisotropy of  growing domains, which are found to exhibit two typical
lengths. Theoretical calculations based on a renormalization group analysis shows that the growth exponent $\alpha_y$ in the shear direction, perpendicular to the flow, is not changed by the applied velocity profile,  while the exponent $\alpha_x$ along the flow direction is increased by $1$ \cite{noisim-1}. Moreover, relevant physical quantities are found to oscillate
on a logarithmic time scale, due to a periodic stretching and breaking-up of domains. The~interest towards the role of shear in phase ordering is still vivid, as
witnessed by recent studies~\cite{naka2021,sara2021}.

In the context of active matter, while efforts have been addressed to investigate the active phase separation in the absence of an external driving {\cite{stenhammar2013,stenhammar2014,speck2014,stenhammar2015,tjhung2018,cates2018}}, much less is known about the dynamic response observed when a shear flow is applied. 
Though it was shown~\cite{caba2018} that in AMB the transition from
homogeneous to bulk phase separation belongs to the same universality class of equilibrium
MB, numerical studies of coarsening in AMB provide different scenarios.
In Refs.~\cite{witt2014,solo2018}, for~example, it was argued that in AMB without shear the size of growing domains is still compatible with a growth exponent $\alpha=1/3$ (i.e., activity has negligible effects on the coarsening dynamics) with a change in the static  phase diagram owing to a pressure jump, even in the case of a flat interface. On~the contrary, very recent simulations~\cite{puri2021-1,puri2021-2} put forward the existence of a late-time growth exponent $\alpha=1/4$, a~result akin to that obtained using particle-based simulations~\cite{stenhammar2013,redner2013}. 

In this work we aim at characterizing, in~the AMB theory,
the shear-induced morphology of growing domains in two spatial dimensions focusing on the role played by the external flow and the activity. 
Our results show that, in~agreement with the AMB without shear, the~activity promotes the formation of isolated drops. This significantly affects the dynamic behavior of the mixture subject to a shear flow. Indeed, unlike the passive counterpart, the~amplitude of the time oscillations of domain sizes diminishes for increasing values of activity, although~their growth exponents are only weakly modified. In~addition, the~probability distribution functions of the lengths of patterns computed along the two spatial directions are found to broaden as the activity~augments. 

The paper is structured as follows. In~Section~\ref{s2} the model is properly defined.  The~numerical results are presented and discussed in Section~\ref{s3}.  The~phenomenology of phase separation is discussed in weak and strong regimes, remarking the differences with respect to the passive model. The~role played by isolated droplets appearing during coarsening is highlighted in connection with the time evolution of the size of domains along the two directions.  Finally, some conclusions are~drawn. 

\section{The~Model}\label{s2}
We consider a system with a scalar 
order parameter $\phi({\bf r},t)$ at position ${\bf r}$ and time $t$.
In the framework of the present model, the~order parameter can be considered as the deviation 
of the density with respect to a reference value.
It satisfies a conserved dynamics given by the following equation
\begin{equation}
\frac {\partial \phi} {\partial t} + {\bf \nabla} \cdot (\phi {\bf v}) = - {\bf \nabla} \cdot {\bf J} .
\label{eqn1}
\end{equation}

Thermal fluctuations are not considered here  as it is usually done when studying phase separation~\cite{BRAY}. At~the l.h.s. a convective term couples $\phi$ to an imposed linear shear flow. This has the form ${\bf v} = \dot\gamma y {\bf e_x}$  where $\dot\gamma$ is the shear rate, $y$ is the coordinate along the
{(shear) $y-$direction} and ${\bf e_x}$ is the unit vector along the
{(flow) $x-$direction}.
{The addition of this term to the original equation of the AMB is the main novelty of this study which allows us to consider the effect of an external velocity profile
on the active coarsening dynamics.}
Hydrodynamic effects are not taken into account since the evolution Equation~(\ref{eqn1}) is not coupled to the  Navier-Stokes equation.
{This is because we are interested in considering only diffusive phase separation. Indeed, to~account for the fluid motion in an active medium, the~dynamics of $\phi$ needs to be coupled to that of a momentum-conserving solvent, whose evolution is governed by the Navier-Stokes equation including an active contribution in the stress tensor~\cite{twm}.}

The current ${\bf J}$ is proportional to the negative gradient of a chemical potential
\begin{equation}
{\bf J} = - \Gamma {\bf \nabla} \mu 
\end{equation}
where the mobility $\Gamma$ is set to unity in the following.
The chemical potential $\mu = \mu_P + \mu_A$
is the sum of passive and active contributions given,
respectively, by~\begin{eqnarray}
\mu_P &=& - a \phi + b \phi^3 - \kappa \nabla^2 \phi \label{mup} , \\
\mu_A &=& \lambda (\nabla \phi)^2 . \label{mua}
\end{eqnarray}

The term (\ref{mup}), where $a, b, \kappa$ are positive and of the order unity, 
corresponds to the passive contribution of the model B, and can be obtained from the derivative of the square gradient $\varphi^4$ free-energy functional.
In the passive case, the system would demix in two coexisting states at binodal densities $\phi_{P1,P2}=\pm \phi_P$ with $\phi_P=\sqrt{a/b}$,
corresponding to the minima of the free-energy density.
The energy cost for the formation of interfaces is proportional to $\kappa$.
The active contribution (\ref{mua}), on~the contrary, cannot be obtained from a proper free energy and is the simplest choice at second order in gradients. It was introduced in the so-called active model B~\cite{witt2014} 
 in order to explicitly break TRS.  For~completeness we note that this term is  similar to the one appearing in {other nonlinear partial differential equations. Among~others we cite, e.g.,}
the Kardar-Parisi-Zhang equation for nonlinear interfacial diffusion~\cite{kpz},
{the Hunter-Saxton equation for nematic liquid crystals
~\cite{hs}, and the Kuramoto-Sivashinsky equation for instabilities
  in laminar flame front~\cite{ks1,ks2}.}
Such a term is controlled by the active parameter $\lambda$ which can be adjusted to go from the passive case ($\lambda=0$) to the active one ($\lambda \neq 0$) with $\lambda \sim O(1)$ \cite{witt2014}. 
We remark that the presence of the $\lambda$ term is such that the system is invariant under the transformation $(\phi, \lambda) \rightarrow -(\phi, \lambda)$. For~this reason we restrict our attention to the case with $\lambda \ge 0$. For~the active model B the binodals $\phi_{1,2}$ ($\phi_1 < 0 < \phi_2$) depend on $\lambda$ and can be calculated by using the method put forward in Ref.~\cite{witt2014} and generalized in Ref.~\cite{solo2018}.
The spinodals do not depend on $\lambda$ \cite{solo2018} and are located at  $\phi_{S1,S2}=\pm \phi_S$ with $\phi_S=\sqrt{a/(3b)}$ as in the model B.  The~state with uniform density is locally unstable between the spinodals while it is metastable between the spinodals and the binodals.
When $\lambda$ increases, the~binodal gets closer to the spinodal on the negative $\phi$ side and stays far on the other~side.

Equation~(\ref{eqn1}) is solved in two dimensions by using a
finite-difference scheme.
The field $\phi$ is discretized on the
nodes ($x_i,y_j$) ($i,j=1,2,...,L$)
of a square lattice with $L \times L$ nodes and mesh size $\Delta x$.
{The time is discretized in 
time steps $\Delta t$ with time values given by
$t^n=n \Delta t$, $n=1,2,3,...$.
Any discretized function $h$ at time $t^n$ 
on a node $(x_i,y_j)$ ($i,j=1,2,...,L$) 
of the lattice is denoted 
by $h(x_i,y_j, t^n)= h_{ij}^n$. At~each time step we update 
$\phi^n \rightarrow \phi^{n+1}$ using an
explicit first-order Euler algorithm for the time derivative~\cite{rogers}
\begin{equation}
\phi^{n+1}=\phi^n 
- \Delta t (\dot\gamma y \partial_x \phi^n
+\partial_{\alpha} J_{\alpha}^{n}) .
\label{adv}
\end{equation}
Central-difference
schemes~\cite{strik} are coded for the spatial derivatives.
The $x$ derivative is given by
\begin{equation}
\partial_{x} h|^n_{ij}=\frac{h^{n}_{(i+1)j}
-h^{n}_{(i-1)j}}{2\Delta x}
\label{xstencil}
\end{equation}
and analogously for the $y$ derivative.
The Laplacian operator appearing in the chemical potential $\mu_P$ is discretized as
\begin{equation}
\nabla^2  h|^n_{ij}=\frac{h^{n}_{(i+1)j}
+h^{n}_{(i-1)j}+h^{n}_{i(j+1)}+h^{n}_{i(j-1)}-4h^{n}_{ij}}{\Delta x^2} 
\label{laplstencil}
\end{equation}}

Periodic boundary conditions (BC) are adopted along the flow direction and
Lees-Edwards BC~\cite{LE} are used in the shear direction.
The latter  take into account the  space shift
$\dot\gamma L \Delta x \Delta t$ occurring in a time step, due to shear,
between the lower and upper row of the lattice.  This is essentially done
by identifying  a point $(x_i,y_1)$ on the lower row of the lattice with
the one placed on the upper row at $(x_i+\dot\gamma L \Delta x \Delta t,y_L)$
($i=1,2,...,L$).

\section{Results}\label{s3}

In the following, all results are obtained with $\Delta x=1$  and $\Delta t=0.001$, values ensuring numerical stability.
We checked that simulations are converged for these values of space and time
discretization units by looking at the behavior of the binodal values and of averaged
domain sizes {for different initial conditions}
(see the following).
{We verified that results are stable upon decreasing the discretization units.
  To this purpose we considered some runs with $\Delta x=0.5$  and $\Delta t=0.0005$, and no significant
differences were found in the binodals as well as in the typical extensions of forming patterns during coarsening.}
The model parameters in Equation~(\ref{eqn1}) are $a=b=1/4$ and $\kappa=1$ while $\lambda$ is varied within the range $[0,3]$.
The shear rate $\dot\gamma$ is changed to access different shear 
regimes in the phase-separation process. To~this purpose we introduce a dimensionless shear rate $\hat{\dot\gamma}=\dot\gamma t_D$ where $t_D$ is the interface diffusion time.
Weak and strong shear regimes are characterized
by values of $\hat{\dot\gamma}$ lower and higher than unit, respectively~\cite{lamu2007}.

\subsection{Planar~Interface}
\begin{figure}[ht]
\includegraphics[width=0.6\columnwidth,angle=0]{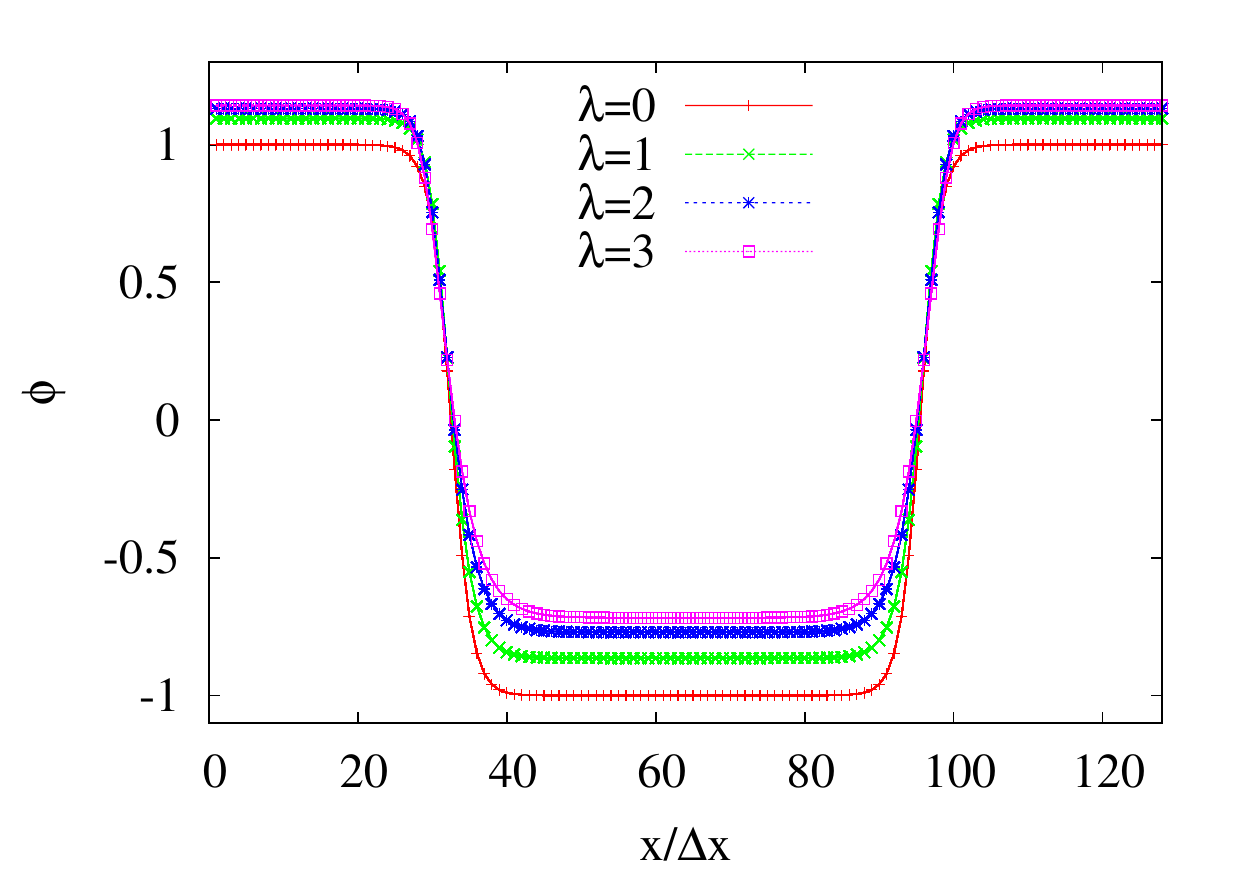}
\caption{Steady density profiles for different values of the parameter  %MDPI: please confirm hyphen should be minus sign. YES
$\lambda$.
\label{fig:profile}
}
\end{figure}

\begin{figure}[ht]
\includegraphics[width=0.7\columnwidth,angle=0]{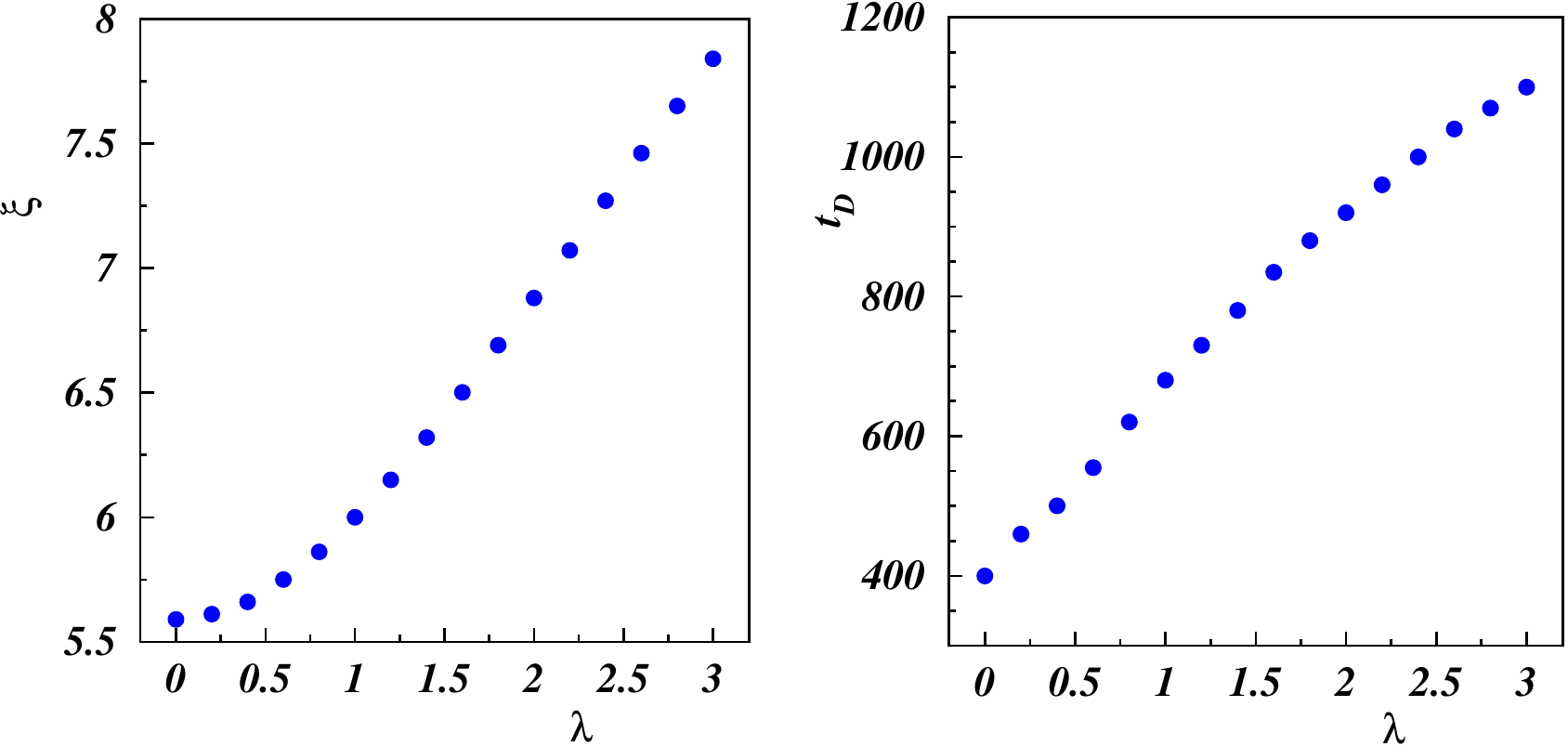}
\caption{Interface width $\xi$ (left panel) and diffusion time $t_D$ (right panel) as functions of the activity $\lambda$.}
\label{fig:fit}
\end{figure}

In the case of the model B one has $t_D=\frac{\xi^3 \phi_P^2}{\Gamma \sigma}$ \cite{fris1997}, where 
$\xi=2 \sqrt{\frac{2 \kappa}{a}}$ is the width
of the planar interface between two coexisting phases 
described  by the function $\phi(x)=\phi_P \tanh{(2x/\xi)}$,
and 
$\sigma=\frac{2}{3}\phi_P^2 \sqrt{2 \kappa a}$ is the interface tension~\cite{BRAY}.
For our choice of the parameters, it is $t_D=384$ in model units.
However, for~the active model B with $\lambda > 0$ there are no explicit expressions 
for $\xi$ and $\sigma$ as well as
for the binodals, though~it can be shown that there is a solution with a planar interface
between the coexisting phases at densities $\phi_{1,2}$ \cite{witt2014}.
For this reason we first compute numerically the binodals $\phi_{1,2}(\lambda)$ by considering the relaxation of a flat interface between two states initially set at the values $\pm \phi_P$.
Once the values  $\phi_{1,2}(\lambda)$ are found, a~sharp
profile between the initial states set at $\phi_{1,2}(\lambda)$ is let to evolve 
to the steady interface.
The stationary profiles are shown in Figure~\ref{fig:profile} for the values $\lambda=0, 1, 2, 3$.
Numerical data are successfully fitted by a kink profile (see the Appendix for further details). 
The values of the interface width $\xi$ obtained from fits for different
values of the activity are
presented in Figure~\ref{fig:fit} and show a linear dependence of $\xi$ on $\lambda$
for $\lambda \gtrsim 1$.
The binodal densities $\phi_1$ and $\phi_2$ increase with
$\lambda$ as $\phi_1$ approaches the spinodal 
$\phi_{S1}=-\phi_S$, being $\phi_S=\sqrt{3}/3$ when $a=b$ (like in our case).
The diffusion time $t_D$ is computed as the time for the initial sharp
profile to relax to the steady interface (see the Appendix  \ref{sA}).
Numerical results give $t_D \simeq 4 \times 10^2$ at $\lambda=0$ in agreement with the
theoretical estimate of the model B. When $\lambda >0$, it is found that $t_D$ increases
with $\lambda$ reaching the highest value $t_D \simeq 1.1 \times 10^3$  when  $\lambda=3$.
The values of $t_D$ are
plotted in the right panel of Figure~\ref{fig:fit} and show that $t_D$ increases
linearly with $\lambda$ up to $\lambda \simeq 1.5$, before slowing down its growth.
In the following we consider a range of values of the shear rate between
$\dot\gamma_w=0.9 \times 10^{-3}$ and $\dot\gamma_s=3.6 \times 10^{-3}$.
The values $\dot\gamma_w$ and $\dot\gamma_s$ are such to access weak 
($\dot\gamma_w t_D(\lambda) \lesssim 1$) 
and strong ($\dot\gamma_s t_D(\lambda) > 1$) shear regimes, respectively,
for all considered values of $\lambda$. 
This guarantees that the shear regime is not affected when changing activity while keeping fixed the strength of applied flow.

\subsection{Phase Separation under Weak and Strong~Shear}
\begin{figure}[ht]
\includegraphics*[width=0.6\columnwidth,angle=0]{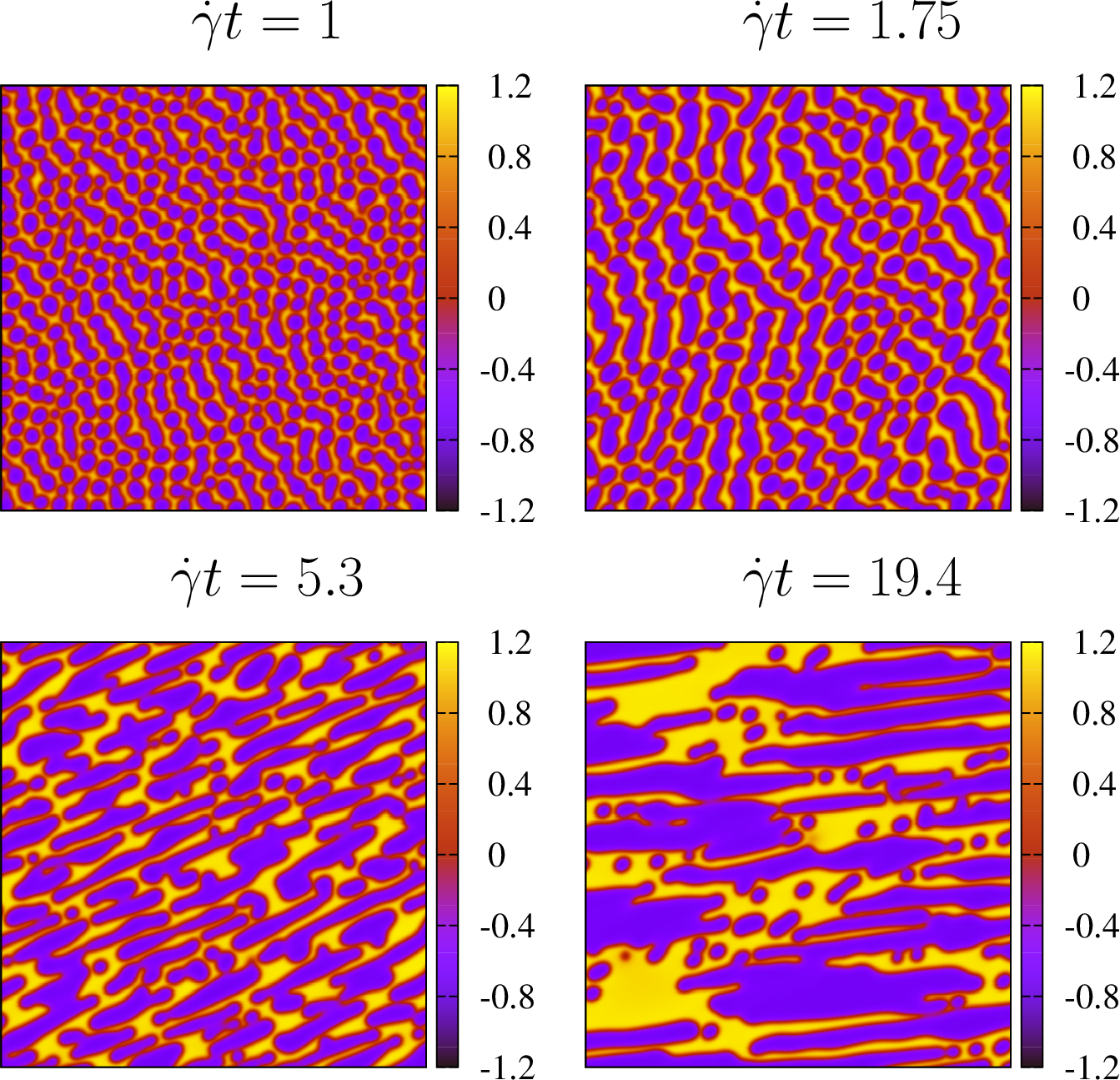}
\caption{Configurations of the system at consecutive times in the weak shear regime ($\dot\gamma t_D=1.0$) with $\lambda=3$. A central portion of size $512 \times 512$ of the whole lattice is
  shown.}  
\label{fig:conf_ws}
\end{figure}

\begin{figure}[ht]
\includegraphics*[width=0.6\columnwidth,angle=0]{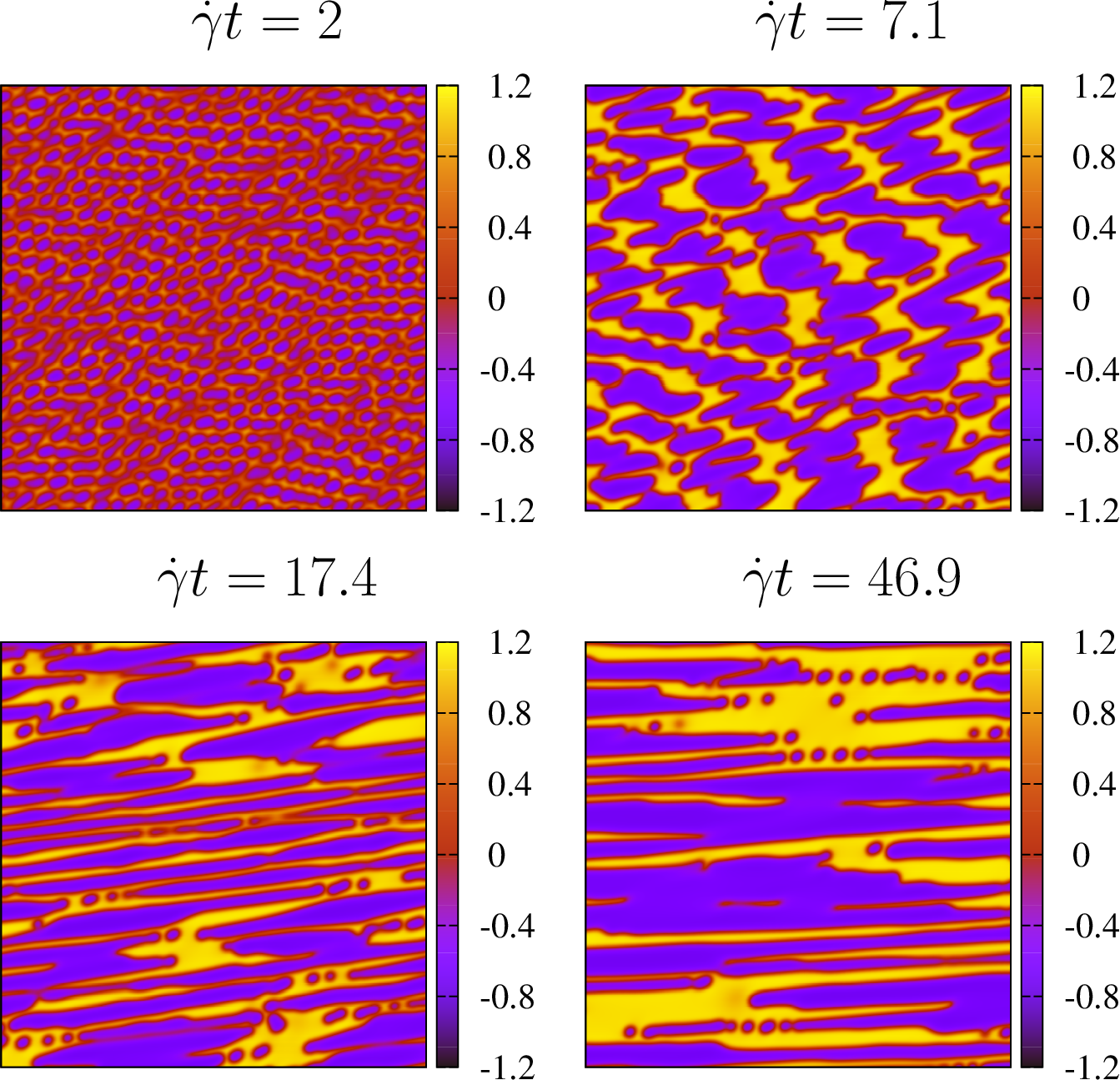}
\caption{Configurations of the system at consecutive times in the strong shear regime ($\dot\gamma t_D=4.0$) with $\lambda=3$. A central portion of size $512 \times 512$ of the whole lattice
  is shown. 
}
\label{fig:conf_ss}
\end{figure}

Now we move on to the study of the phase separation under an external shear flow by varying the 
activity parameter $\lambda$ for $\dot\gamma=\dot\gamma_w, \dot\gamma_s$. 
We consider a system initially prepared in a symmetric 
disordered state with $\varphi({\bf r},0)=\omega$ where $\omega$ is a random number in the
range $[-0.01,0.01]$. This state corresponds to a critical composition of the system.
The size of the lattice is $L=1024$ and measures are averaged over five independent runs. The different values of the dimensionless shear rate $\hat{\dot\gamma}$ influence 
the initial morphology of the forming domains. This can be seen in 
Figures~\ref{fig:conf_ws} and \ref{fig:conf_ss} where snapshots of systems at consecutive times are
shown for the cases at weak ($\dot\gamma t_D=1.0$)
and strong shear ($\dot\gamma t_D=4.0$), respectively, with~$\lambda=3$. 
At weak shear, domains form and grow while the density $\phi$ attains 
the steady binodal values before shear effects come into play 
at values of strain $\dot\gamma t > 1$. This occurs since the relaxation time scale
$t_D(\lambda)$ is smaller than the shear characteristic time $1/\dot\gamma$. Once 
isolated drops at density $\phi_1$ are produced ($\dot\gamma t=1$), 
the applied flow advects them along the $x-$direction and favours their merging, thus causing the formation of elongated domains along the $y-$direction 
($\dot\gamma t=1.75$). Afterwards the shear stretches these structures which are tilted as well, 
being characterized by different thicknesses. 
{Isolated droplets at density 
$\phi_1$ can be observed inside domains at density $\phi_2$ ($\dot\gamma t=5.3$)}. 
Later on, the~shear further deforms and tilts the domains, which may eventually break up. After~the overstretching, domains 
retract forming $\phi_2$ phases
with a larger thickness where isolated droplets of the other phase get trapped 
($\dot\gamma t=19.4$). Such stretching and bursting persist
though this periodic behavior cannot be observed over very long periods of time due to the finite size of the~system.

\begin{figure}[ht]
\includegraphics*[width=0.6\columnwidth,angle=0]{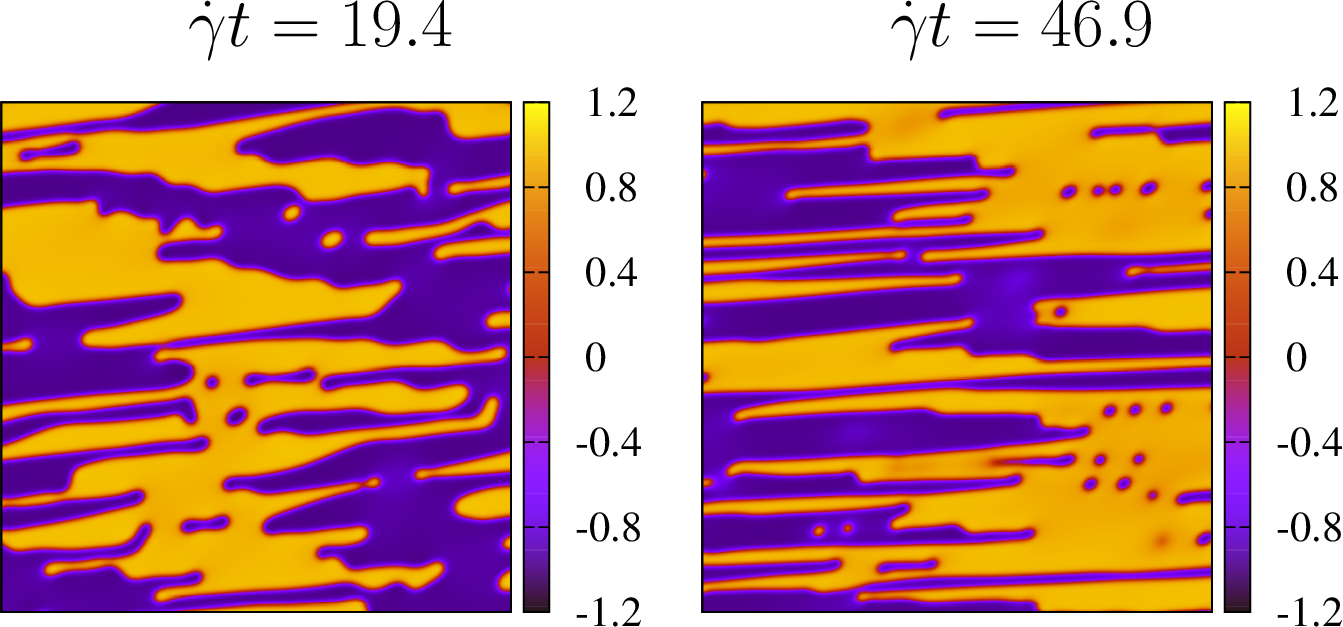}
\caption{Configurations of the system in the weak (left panel, $\dot\gamma t_D=0.4$) 
and strong (right panel, $\dot\gamma t_D=1.4$) shear regimes with $\lambda=0$. 
Snapshots are taken at the latest times of Figures~\ref{fig:conf_ws} and \ref{fig:conf_ss}.
A central portion of size $512 \times 512$ of the whole lattice is shown. }
\label{fig:conf_nolambda}
\end{figure}

In the strong shear regime, the initial morphology is different with respect to the previous case, since the interface diffusion time is larger than the inverse of the shear rate. 
Indeed, while domains grow (as quantified later) and start to be deformed by the flow,  the~two phases are still far from the steady binodal densities (see the panel at $\dot\gamma t =2$ in Figure~\ref{fig:conf_ss}), an~effect promoting an initial 
thinning of domains along the shear direction. Afterwards, the~elongations and ruptures of domains previously described take place once again,
a phenomenon also observed in simulations of passive model B with strong shear~\cite{noisim-1,noisim-2}.
However, unlike such cases, the~main feature here is that, {\it under shear}, isolated droplets of the $\phi_1$ phase survive and are dispersed in the $\phi_2$ matrix (see in particular the last snapshots of Figures~\ref{fig:conf_ws} and \ref{fig:conf_ss}). This is not the case if $\lambda=0$, as~demonstrated in Figure~\ref{fig:conf_nolambda} 
where instantaneous configurations of the passive case taken at equal times are shown.

A scenario akin to that just described in the AMB occurs
when considering sheared binary mixtures with surface diffusion~\cite{surfdiff}. 
However, in~the present work the physical explanation is different.
The fraction $\beta$ of the $\phi_1$ phase is larger than the one  
of the $\phi_2$ phase when $\lambda > 0$ despite the fact that
we consider a critical system with a 
symmetric initial composition such that $<\phi>=0$ (the symbol
$<...>$ denotes an average over the lattice). Since the dynamics described
by Equation~(\ref{eqn1}) is conserved, the~average value of the order parameter $\phi$ is preserved
and it has to be 
$\beta \phi_1 + (1-\beta)\phi_2=0$.
In a passive mixture the values of the binodals $\phi_1, \phi_2$
are symmetric with respect to the value
$\phi=0$ and it results $\beta=0.5$.
In the AMB the binodals are not symmetric anymore with respect to the value
$\phi=0$ and one has $\beta>0.5$, since $|\phi_1|<\phi_2$ when $\lambda>0$.  
As a result, the~$\phi_1$ phase is more abundant than the $\phi_2$ one.
In our study it is $\phi_1 \simeq -0.71$ and $\phi_2 \simeq 1.15$ when $\lambda=3$
so that $\beta \simeq 0.62$. 
Therefore, in~AMB, it is the activity that produces an effective off-symmetric mixture though the initial 
state is not, in~agreement with previous studies of  AMB~\cite{witt2014}.
Such off-symmetric mixture consists of droplets of the majority ($\phi_1$) phase dispersed in the minority ($\phi_2$) one, a~situation persisting under shear.
We finally note that a different morphology would be observed in the coarsening of off-symmetric 
passive mixtures under shear with a similar ratio between the two phases~\cite{offcrit}.
In this case, at~low shear rate small droplets evaporate due to Ostwald ripening, while for strong flows it is
the minority phase to be dispersed in a large number of~droplets.

\subsection{Domain~Size}
\begin{figure}[ht]
\includegraphics*[width=0.99\columnwidth,angle=0]{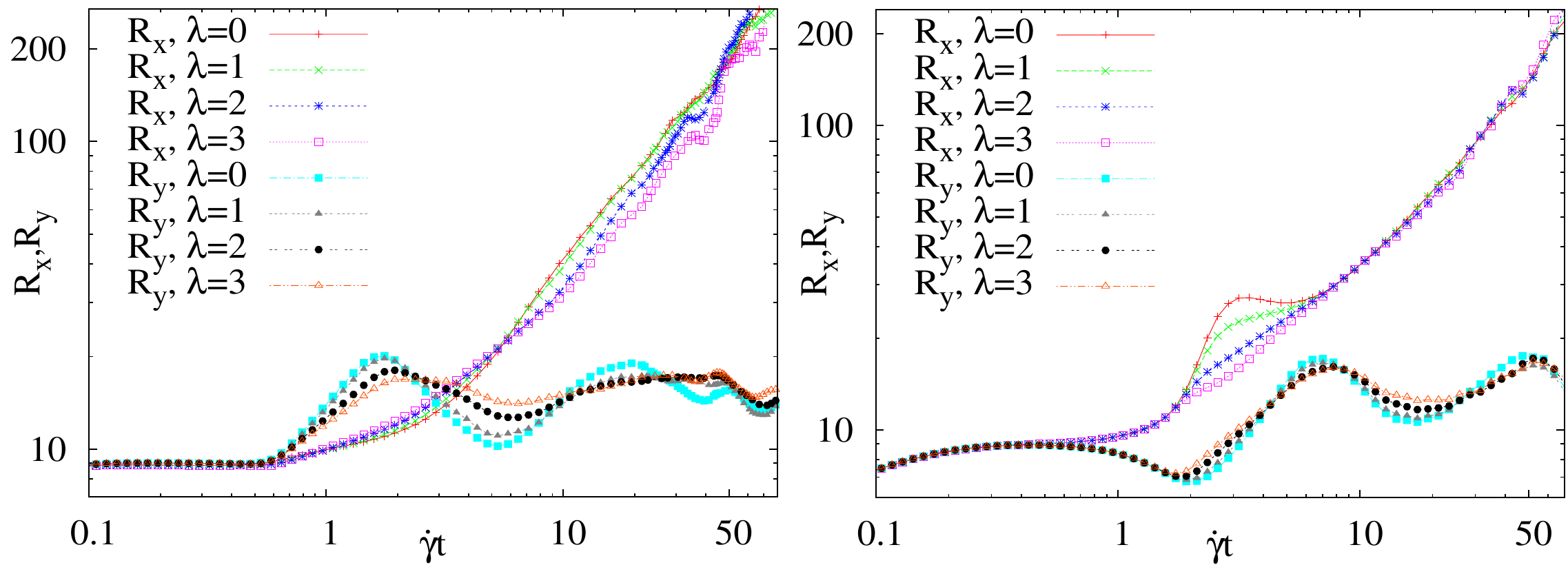}
\caption{Average measures of domains along the flow ($R_x$) and the shear ($R_y$) directions 
as functions of time at weak (left panel) and strong (right panel) shear
for different values of $\lambda$.
\label{fig:rad}
}
\end{figure}

The coarsening dynamics can be investigated by looking at evolution of the typical measures of domains.
The sizes $R_x$ and $R_y$ along the flow and the shear directions, respectively,
are shown in Figure~\ref{fig:rad} and are
 computed as the inverse of the first moments of the structure factor
\begin{equation}
R_{x,y}(t)=\pi \frac{\int d{\bf k} C({\bf k},t)}{\int d{\bf k} |k_{x,y}| C({\bf k},t)} .
\end{equation}
$C({\bf k},t)=\langle \phi({\bf k},t)\phi({\bf -k},t)\rangle$ is the structure factor averaged over different realizations of the system and 
$\phi({\bf k},t)$ is the Fourier transform of the order~parameter.

At weak shear, it results that $R_x \simeq R_y$ until $\dot\gamma t \lesssim 1$, then $R_y$
grows initially faster than $R_x$ due to the shear-induced dragging of droplets along the
flow direction. 
The typical size along the shear direction attains a maximum at $\dot\gamma t \simeq 1.75$ and then decreases, while the growth rate along the flow direction increases due to shear stretching. 
The quantity $R_y$ oscillates on a logarithmic time scale between minimum and maximum values corresponding to elongated and broken domains, respectively. The~amplitude of such oscillations diminishes when increasing the activity parameter $\lambda$ and is related to the existence of small droplets that cannot grow any further. The~radii $R_x$ for different values of $\lambda$ show a similar trend compatible with an asymptotic power-law growth
$R_x \sim A(\lambda) t^{\alpha_x}$.
Fitting the power law for $R_x$ to numerical data shows that the amplitudes $A$ decrease with the activity as illustrated in Table~\ref{table}, while the exponent $\alpha_x \simeq 1$ does not seem to be affected by $\lambda$.
\begin{table}
\begin{center}
\begin{tabular}{c|l}
$\lambda$&A\\\hline
0&4.40\\
1&2.75\\
2&0.99\\
3&0.82
\end{tabular}
\end{center}
\caption{Fitted values of the amplitudes $A$ for different values of the activity $\lambda$
in the weak shear regime. 
\label{table}
}
\end{table}

%\begin{specialtable}[h]
%\caption{Fitted values of the amplitudes $A$ for different values of the activity $\lambda$
%in the weak shear regime. \label{table1}}
%\setlength{\cellWidtha}{\columnwidth/2-2\tabcolsep-0in}
%\setlength{\cellWidtha}{\columnwidth/2-2\tabcolsep-0in}
%\scalebox{1}[1]{\begin{tabularx}{\columnwidth}{
%>{\PreserveBackslash\centering}m{\cellWidtha}
%>{\PreserveBackslash\centering}m{\cellWidthb}}
%\toprule
%\boldmath$\lambda$&\textbf{A}\\\midrule
%0&4.40\\
%1&2.75\\
%2&0.99\\
%3&0.82\\
%\bottomrule
%\end{tabularx}}
%\end{specialtable}
Due to the limited size of the simulated system and the superimposed oscillations, it is
difficult to estimate the growth exponent $\alpha_y$ along the shear direction
although, at~$\dot\gamma t \gtrsim 5$, one observes a regime with 
$\alpha_x-\alpha_y \simeq 1$.  This result is due to the
advection of domains by the applied flow, and~has been observed in the passive model B under shear as well~\cite{noisim-1,noisim-2}.

At increasing values of shear rate, the~growth along the flow direction proceeds faster than that in the shear direction, with~a local maximum of $R_x$ at $\dot\gamma t \simeq 3$ corresponding to a minimum for $R_y$, since domains are strongly deformed by the shear. Note that the height of the maximum of $R_x$ is reduced by increasing the activity, a~feature associated to the presence of droplets which can be deformed only slightly by the flow. When $\dot\gamma t \gtrsim 7$,  $R_x$ shows a power-law growth with an exponent $\alpha_x \simeq 1.1$ not depending on $\lambda$, while
the size $R_y$ exhibits a periodic behavior on a logarithmic time scale with amplitude that shrinks with $\lambda$. 
By using power counting, it might be expected that the net effect of the
convective term is to increase the growth exponent of the unsheared system by $1$  in the flow
direction. However, as~previously discussed, in~AMB with no external flow the value of the growth exponent has not been definitely determined yet. Our results seem to point towards a picture in which the activity has mild effect on the coarsening dynamics, as~in the unsheared case~\cite{witt2014}, since we find that $\alpha_x\simeq 1$ as in the passive case and $\alpha_x-\alpha_y\simeq 1$. 
Additional simulations run on systems of size $L=512$ do not provide further insights for evaluating $\alpha_x$ and $\alpha_y$ as well as for attempting a finite-size scaling analysis. Indeed, a~more accurate estimate of the growth exponents would likely require much larger systems, thus dramatically increasing 
the computational resources necessary to simulate their late-time~dynamics.  

To elucidate the role of flow strength, simulations with shear rate 
$\dot\gamma = 2.8 \times 10^{-3}$ are also considered. This value is such that $\dot\gamma t_D(\lambda) \gtrsim 1.1$ for $0 \leq \lambda \leq 3$ so that the system is in the strong regime. The~time behavior of $R_x$ and $R_y$ is similar to the one shown in Figure~\ref{fig:rad}, with~no significant effect on the growth exponents.
We only observe a reduction in the amplitudes of oscillations
at the smallest values of activity along the $x$-direction, since domains are less affected by flow.  
Being less deformed, they can grow along the shear direction, a~result witnessed by wider amplitudes of $R_y$ with respect to the case with $\dot\gamma=3.6 \times 10^{-3}$.
These effects reduce when $\lambda$ increases, since, as~previously discussed, the~dynamics is deeply modified by the presence of droplets for the highest value of~activity.

\begin{figure}[ht]
\includegraphics*[width=0.99\columnwidth,angle=0]{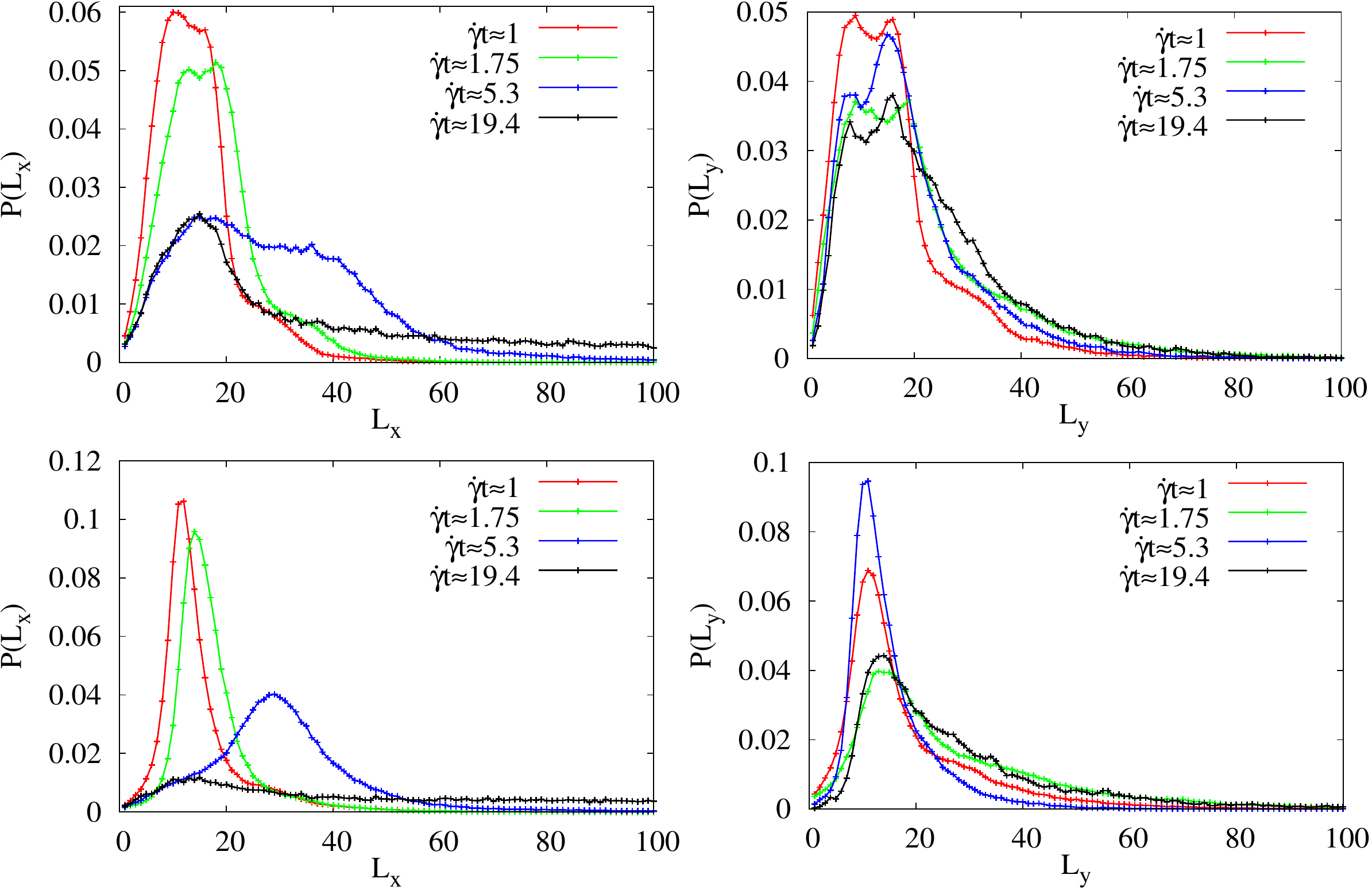}
\caption{Probability distribution functions $P$ of domains of length $L_x$ (left panels) 
and $L_y$ (right panels) with $\lambda=3$ (upper row) and $\lambda=0$ (lower row)
in the weak shear regime. 
\label{fig:histo_ws}
}
\end{figure}

\begin{figure}[ht]
\includegraphics*[width=0.99\columnwidth,angle=0]{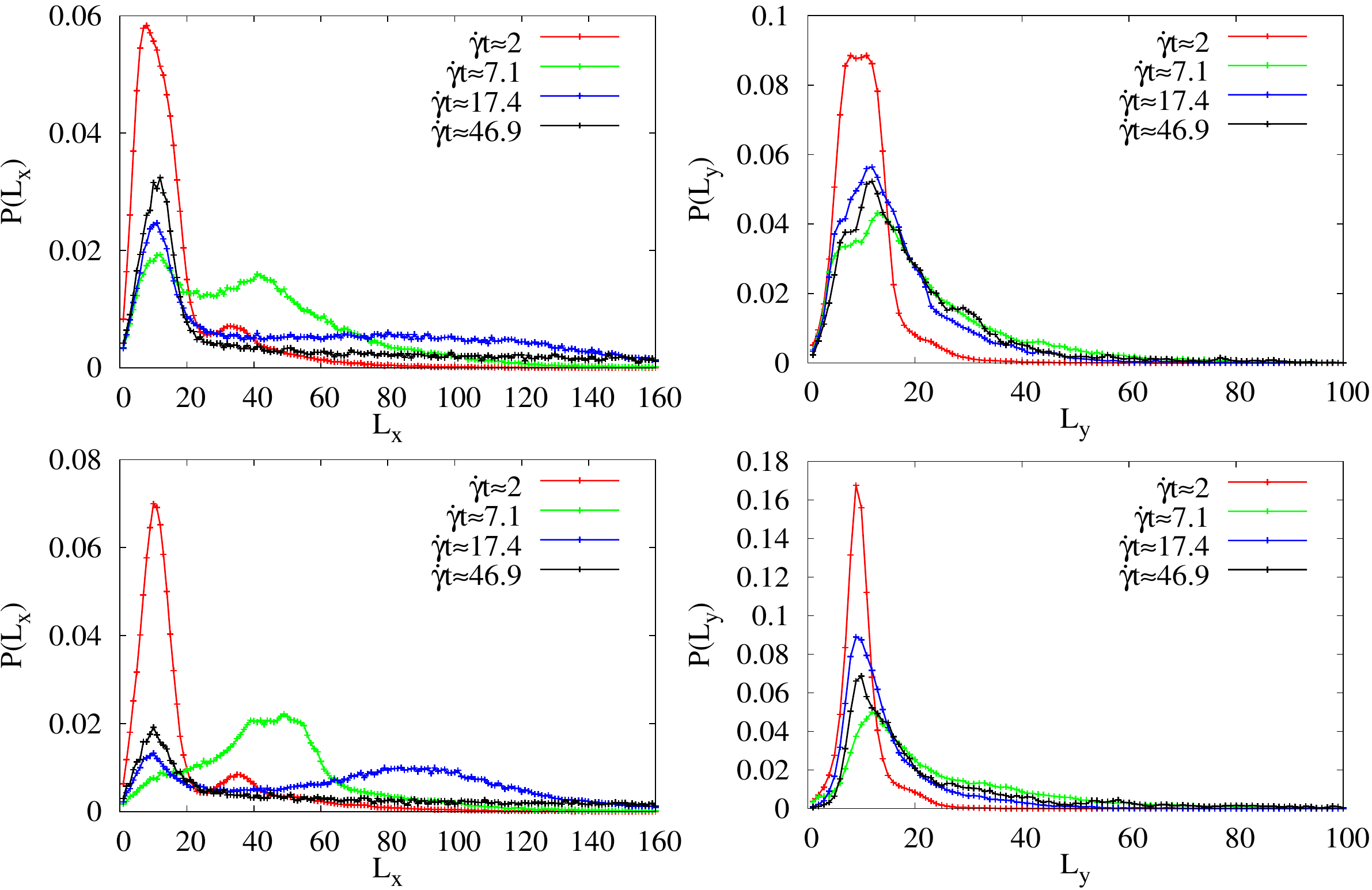}
\caption{Probability distribution functions $P$ of domains of length $L_x$ (left panels) 
and $L_y$ (right panels) with $\lambda=3$ (upper row) and $\lambda=0$ (lower row) in the strong shear regime. 
\label{fig:histo_ss}
}
\end{figure}

The size distribution of domains can be analyzed by calculating the normalized probability distributions $P(L_{x,y})$, having domains of lengths $L_x$ and $L_y$ along the flow and the shear directions, respectively. 
By moving along all rows (the flow direction) of the computational domain, 
$L_x$ is computed as the size of unidimensional domains with equal composition (same sign of the density $\phi$). From~the registered values of $L_x$, the~function $P(L_x)$ is derived.
The same procedure is adopted along the columns (the shear direction) of the lattice to determine $L_y$ and then $P(L_y)$.
In Figures~\ref{fig:histo_ws} and \ref{fig:histo_ss} we plot $P(L_x)$ and $P(L_y)$ for $\lambda=0, 3$ in the weak and strong shear regime. 
At weak shear (see Figure~\ref{fig:histo_ws}), $P(L_x)$ and $P(L_y)$ show two peaks in the active case and a single one in the passive situation. The~peaks at the smaller values of $L_x$ and $L_y$ (with $L_x \simeq L_y \simeq 8$) correspond to the presence of isolated droplets, while the other peaks indicate elongated domains. In~particular, $P(L_x)$ broadens when going from the maximum of $R_y$ at $\dot\gamma t \simeq 1.75$ to the minimum at $\dot\gamma t \simeq 5.3$ and, then, shrinks when $R_y$ grows again.  In~the passive counterpart, the~distribution functions are less broad. When looking at the shear direction, 
in the active case the peak of
$P(L_y)$ at the larger value of $L_y$ oscillates in height, with~a local maximum attained when $R_y$ is
at the minimum ($\dot\gamma t \simeq 5.3$). Later $P(L_y)$ broadens. In~the passive case the distribution $P(L_y)$
has a single peak (narrower than in the active case) that oscillates in height, following the cyclic dynamics of elongations and~ruptures.
\unskip

Moving to the strong shear regime (Figure~\ref{fig:histo_ss}), we observe that
$P(L_x)$ is characterized by the presence of two peaks both in the passive and in the active case,  roughly at the same positions. When $\lambda=3$, the~peak at $L_x \simeq 10$ is always the prevailing one, owing to the presence of droplets spanning the system. 
The value of $L_x$, corresponding to the second peak, increases in time attaining its maximum at
$\dot\gamma t \simeq 17.4$, when domains are highly stretched ($R_y$ is at its minimum). The~subsequent
breaking-up of domains determines the shift in position of the second peak to smaller values. 
Moreover, it can be seen that the distributions are broader in the active case.
This feature also holds for $P(L_y)$, whose main peak oscillates in height during the
time evolution. We note here that in the active system the first peak of $P(L_y)$ 
at the smaller value of $L_y$, corresponding to isolated droplets, is less pronounced with respect to that of  the weak shear regime. This can be attributed to the fact that, as~previously discussed, initially forming droplets are fused by shear before growing in~size.  

\section{Conclusions}\label{s4}
To summarize, we have investigated the phase separation of an active binary mixture subject to an applied shear flow. For~this purpose we numerically solved the phenomenological equation of the active model B~\cite{witt2014}, supplemented by a convective term that couples the order parameter to the external velocity field. 
The initial morphology depends on the strength of shear. Later, growing domains are elongated, tilted, and~burst by the flow. This is reflected in the typical sizes of domains $R_x$ and $R_y$ along the two spatial dimensions, which appear to be modulated by oscillations on a logarithmic time scale. However, the~presence of activity is such that the fraction of one phase is larger than the other one, despite the initial symmetric composition. This induces the presence of droplets of the more abundant phase which span the system and are responsible for the observed reduction of the amplitudes of the oscillations of $R_x$ and $R_y$ when increasing the activity parameter. Though~the limited size of the simulated system does not allow an estimate of the growth exponents
$\alpha_x$ and $\alpha_y$ along the flow and the shear directions, respectively, we find that $\alpha_x-\alpha_y \simeq 1$ as in the passive case~\cite{noisim-1,noisim-2}.
The combined effect of activity and shear on the overall morphology  
has been studied by considering the probability distribution functions (PDFs) of the size of patterns along the two spatial directions. Our simulations suggest that such PDFs are characterized by a width that broadens with~activity. 

We hope that our results may stimulate further research on this system. It would be of interest, for~example, understanding the role played by the shear in active binary mixtures where all terms breaking time-reversal symmetry to leading order in $\nabla$ and $\phi$ are included~\cite{tjhung2018}, as~well as how thermal noise is expected to impact on morphology and growth dynamics in the presence of an external~driving.

\section*{Acknowledgements}
The authors wish to thank M. E. Cates for useful comments. A. T. acknowledges funding from the European Research Council under the European Union's Horizon 2020 Framework Programme (No. FP/2014-2020) ERC Grant Agreement No.739964 (COPMAT).

\section*{Appendix}\label{sA}
Here we detail how the values of $\xi$ and $t_D$ in Figure~\ref{fig:fit}
are measured.
The interface width $\xi$ is obtained by fitting the steady interface profile along the $x$-direction with the following functional form
\begin{equation}
\phi(x,\lambda)=\phi_1(\lambda)+\frac{(\phi_2(\lambda)-\phi_1(\lambda))}{2}
\left[1+\tanh{\left(\frac{2x}{\xi}\right)}\right],
\end{equation}
where $\phi_{1,2}(\lambda)$ are the binodals and $\xi$ is the only fit~parameter.
 
The values of $t_D$ are evaluated by computing the ratio
\begin{equation}\label{eq_t_d}
\Lambda(t)=\frac{\sum_i |\phi_i(t)-\phi_i(t-t_1)|}{\sum_i|\phi_i(t-t_1)|},
\end{equation}
over time, where $\phi_i$ is the density along a unidimensional profile initialized between the binodals $\phi_1$ and $\phi_2$ and $i$ is a discrete lattice index running over the whole planar interface ($i=1,...,L$). 
This quantity provides an estimate of the relative difference between two profiles taken at a time interval equal to $t_1$. The~relaxation time $t_D$
is the time such that $\Lambda(t_D)<10^{-2}$.
A reasonable estimate of $t_D$ is obtained by using $t_1=50$. 
Although other values of $t_1$ can be adopted, our choice represents a reasonable compromise between the requirement $t_1 \ll t_D$ to resolve $t_D$ and the need of having a 
sufficiently large time interval to avoid indistinguishable profiles at times
$t-t_1$ and $t$.

\end{document}